\documentclass[%
reprint,
superscriptaddress,
amsmath,
amssymb,
aps,
prb,
]{revtex4-1}

\usepackage{graphicx}
\usepackage{subfigure}
\usepackage{dcolumn}
\usepackage{bm}
\usepackage{amsmath}
\usepackage[colorlinks=true,citecolor=blue, urlcolor = blue, linkcolor= blue]{hyperref}
\usepackage{epstopdf}
\usepackage{color,soul}
\usepackage[toc,page]{appendix}

\makeatletter
\AtBeginDocument{\@ifpackageloaded{natis}{\ifNAT@numbers\if@filesw\immediate\write\@auxout{\string\global\string\NAT@numberstrue}\fi\fi}{}}
\makeatother

\makeatletter
\AtBeginDocument{\@ifpackageloaded{natbib}{\ifNAT@numbers\if@filesw\immediate\write\@auxout{\string\global\string\NAT@numberstrue}\fi\fi}{}}
\makeatother

\begin{document}

\title{Atomistic structural mechanism for the glass transition: Entropic contribution}

\author{Dong Han}
\affiliation{State Key Laboratory of Nonlinear Mechanics, Institute of Mechanics, Chinese Academy of Sciences, Beijing 100190, China}
\affiliation{School of Engineering Science, University of Chinese Academy of Sciences, Beijing 100049, China}

\author{Dan Wei}
\affiliation{State Key Laboratory of Nonlinear Mechanics, Institute of Mechanics, Chinese Academy of Sciences, Beijing 100190, China}
\affiliation{School of Engineering Science, University of Chinese Academy of Sciences, Beijing 100049, China}

\author{Jie Yang}
\affiliation{State Key Laboratory of Nonlinear Mechanics, Institute of Mechanics, Chinese Academy of Sciences, Beijing 100190, China}
\affiliation{School of Engineering Science, University of Chinese Academy of Sciences, Beijing 100049, China}

\author{Hui-Ling Li}
\affiliation{State Key Laboratory of Nonlinear Mechanics, Institute of Mechanics, Chinese Academy of Sciences, Beijing 100190, China}
\affiliation{School of Engineering Science, University of Chinese Academy of Sciences, Beijing 100049, China}

\author{Min-Qiang Jiang}
\affiliation{State Key Laboratory of Nonlinear Mechanics, Institute of Mechanics, Chinese Academy of Sciences, Beijing 100190, China}
\affiliation{School of Engineering Science, University of Chinese Academy of Sciences, Beijing 100049, China}

\author{Yun-Jiang Wang}
\email{yjwang@imech.ac.cn}
\affiliation{State Key Laboratory of Nonlinear Mechanics, Institute of Mechanics, Chinese Academy of Sciences, Beijing 100190, China}
\affiliation{School of Engineering Science, University of Chinese Academy of Sciences, Beijing 100049, China}

\author{Lan-Hong Dai}
\email{lhdai@lnm.imech.ac.cn}
\affiliation{State Key Laboratory of Nonlinear Mechanics, Institute of Mechanics, Chinese Academy of Sciences, Beijing 100190, China}
\affiliation{School of Engineering Science, University of Chinese Academy of Sciences, Beijing 100049, China}

\author{Alessio Zaccone}
\email{az302@cam.ac.uk}
\affiliation{Department of Physics ¡°A. Pontremoli¡±, University of Milan, via Celoria 16, 20133 Milan, Italy}
\affiliation{Cavendish Laboratory, University of Cambridge, JJ Thomson Avenue, Cambridge, CB3 0HE United Kingdom}
\affiliation{Statistical Physics Group, Department of Chemical Engineering and Biotechnology, University of Cambridge, Cambridge, CB3 0AS, United Kingdom}

\date{\today}

\begin{abstract}
A popular Adam--Gibbs scenario has suggested that the excess entropy of glass and liquid over crystal dominates the dynamical arrest at the glass transition with exclusive contribution from configurational entropy over vibrational entropy. However, an intuitive structural rationale for the emergence of frozen dynamics in relation to entropy is still lacking. Here we study these issues by atomistically simulating the vibrational, configurational, as well as total entropy of a model glass former over their crystalline counterparts for the entire temperature range spanning from glass to liquid. Besides confirming the Adam--Gibbs entropy scenario, the concept of Shannon information entropy is introduced to characterize the diversity of atomic-level structures, which undergoes a striking variation across the glass transition, and explains the change found in the excess configurational entropy. Hence, the hidden structural mechanism underlying the entropic kink at the transition is revealed in terms of proliferation of certain atomic structures with a higher degree of centrosymmetry, which are more rigid and possess less nonaffine softening modes. In turn, the proliferation of these centrosymmetric (rigid) structures leads to the freezing-in of the dynamics beyond which further structural rearrangements become highly unfavourable, thus explaining the kink in the configurational entropy at the transition.
\end{abstract}

\maketitle

\section{Introduction}

The glass transition is generally regarded as the phenomenon in which a viscous liquid circumvents crystallization and evolves continuously into a disordered solid state directly during fast cooling \cite{Angell1995,debenedetti2001supercooled,Dyre2006,Langer2014,Royall2015}. It is a typical example of the falling-out-of-equilibrium phenomenon that occurs for almost any system the relaxation time of which surpasses laboratory time scales \cite{Dyre2008}. Gibbs and DiMarzio~\cite{gibbs1958nature} suggested that the excess entropy of glass and liquid over crystal originates entirely from the configurational entropy which governs the relaxation timescale. This formulation lays a robust foundation for the Adam--Gibbs relationship, which provides a connection between dynamics and thermodynamics of glass transition, i.e., time and entropy \cite{adam1965temperature,Martinez2001,dyre2009brief,Dyre2018}.

Also in the Potential Energy Landscape (PEL) picture~\cite{debenedetti2001supercooled,Sastry1998,Fan2017}, it is assumed that the vibrational entropy is in a relation of linear response with temperature  \cite{goldstein1976viscous} and plays a minor role compared to configurational entropy. A recent  simulation also revealed that the ideal glass state is not only vibrational \cite{Ozawa2018a}. To validate the entropic scenario, configurational and vibrational contributions to the excess entropy have been evaluated for molecular and network glasses \cite{gujrati1980viscous,johari2000contributions}, as well as computer Lennard-Jones liquids \cite{Sciortino1999,Ozawa2018}; however, there are only very few reports on metallic glass-forming systems available in the literature.

It is the lack of thermodynamic stability of supercooled liquids against crystallization in experiments that hinders separating the vibrational and configurational entropy across the glass transition. Recently, Smith \textit{et al}. have successfully obtained the phonon density of states (DOS) thanks to the advances in neutron flux and instrument efficiency for inelastic neutron scattering (INS), which enables capturing of vibrational states in a very short time window feasible above $T_{\rm g}$~\cite{smith2017separating}. These experiments suggest that vibrational entropy is indeed trivial, or featureless, against its configurational counterpart.

Nevertheless, an entropic picture including atomistic information in the entire temperature space is missing. As a result, a question naturally arises about which atomic-level structures change the most across the transition \cite{Royall2015,Tong2018,Tanaka2019}. While there exist some hints which may be recognized as structural signatures of glass transition \cite{Shen2009,ding2012short,hu2015five,Schoenholz2016}, an explicit link between structural mechanism and configurational entropy for the dynamical arrest is missing \cite{Royall2008,Royall2015,Tong2018,Tanaka2019}.

Here we fill this gap by disentangling the specific contributions of vibrational and configurational entropy across the glass transition and by relating the configurational entropy to the distribution of atomic-level structures. Upon introducing a Shannon information entropy measure of local structural diversity, we rationalize the structural mechanism for the glass transition which is responsible for the variation in configurational entropy. The resulting scenario also offers the unprecedented perspective of linking entropy, structure and mechanical properties into a single unifying framework.

\section{Simulation details}

The molecular dynamics (MD) simulations are performed by LAMMPS \cite{plimpton1995fast} on a prototypical Cu$_{50}$Zr$_{50}$ glass-forming liquid which has been widely studied in simulations. The force field is described by a Finnis-Sinclair type embedded-atom method (EAM) potential \cite{Mendelev2009}. A model simulation box containing 10,976 atoms with dimensions 60 $\AA$ $\times$ 60 $\AA$ $\times$ 60 $\AA$ is used for estimating the phonon density of states. A bigger simulation box with 31,250 atoms is adopted for studying the structural motifs and their statistical occurrence. The glass sample is prepared by quenching an equilibrium liquid at 2000 K to 0 K with a cooling rate of 10$^{10}$ K/s. A constant temperature, pressure and atom number ensemble is used for both cooling and heating. The phonon is obtained by diagonalization of the Hessian matrix. Through Intel Math Kernel Library and LAPACK, we diagonalize the Hessian matrix based on the EAM formulation to obtain the vibrational normal modes \cite{Widmer-Cooper2008,Shintani2008,Ding2016b}. Then the phonon DOS is obtained in a standard way as $D(\omega)=\frac{1}{3 N-3} \sum_{l=1}^{3 N-3} \delta\left(\omega-\omega_{l}\right)$, with $N$ being the number of atoms, $l$ the number of vibrational modes, and $\omega$ the eigenfrequency. Finally, the local structure motifs of inherent structures are categorized by the Voronoi tessellation. A standard four digit descriptor $\left\langle{{n_3},{n_4},{n_5},{n_6}}\right\rangle$ is used to label a Voronoi polyhedron, where $n_i$ is the number of facets with $i$ edges around a central atom \cite{Sheng2006,Cheng2010}.

\section{Results and discussion}

To quantitatively decouple the specific roles of vibrational entropy $S_{\rm vib}$ and of configurational entropy $S_{\rm conf}$ across the glass transition, we treat the total entropy as ${S_{\rm{tot}}} = S_{\rm vib} + S_{\rm conf}$, which can be obtained by thermodynamic integration after heating a glass to liquid, i.e.,
\begin{equation}\label{eq1}
{S_{\rm{tot}}} = \int_0^T {\frac{{{\rm{d}}Q}}{T}}  = \int_0^T {\frac{{{\rm{d}}U}}{T}} \left( {P = 0} \right).
\end{equation}
Here $T$ is temperature, $Q$ is the absorption heat, $U$ is internal energy, and $P$ is pressure.

\begin{figure}
\centering
 {\includegraphics[width=0.48\textwidth] {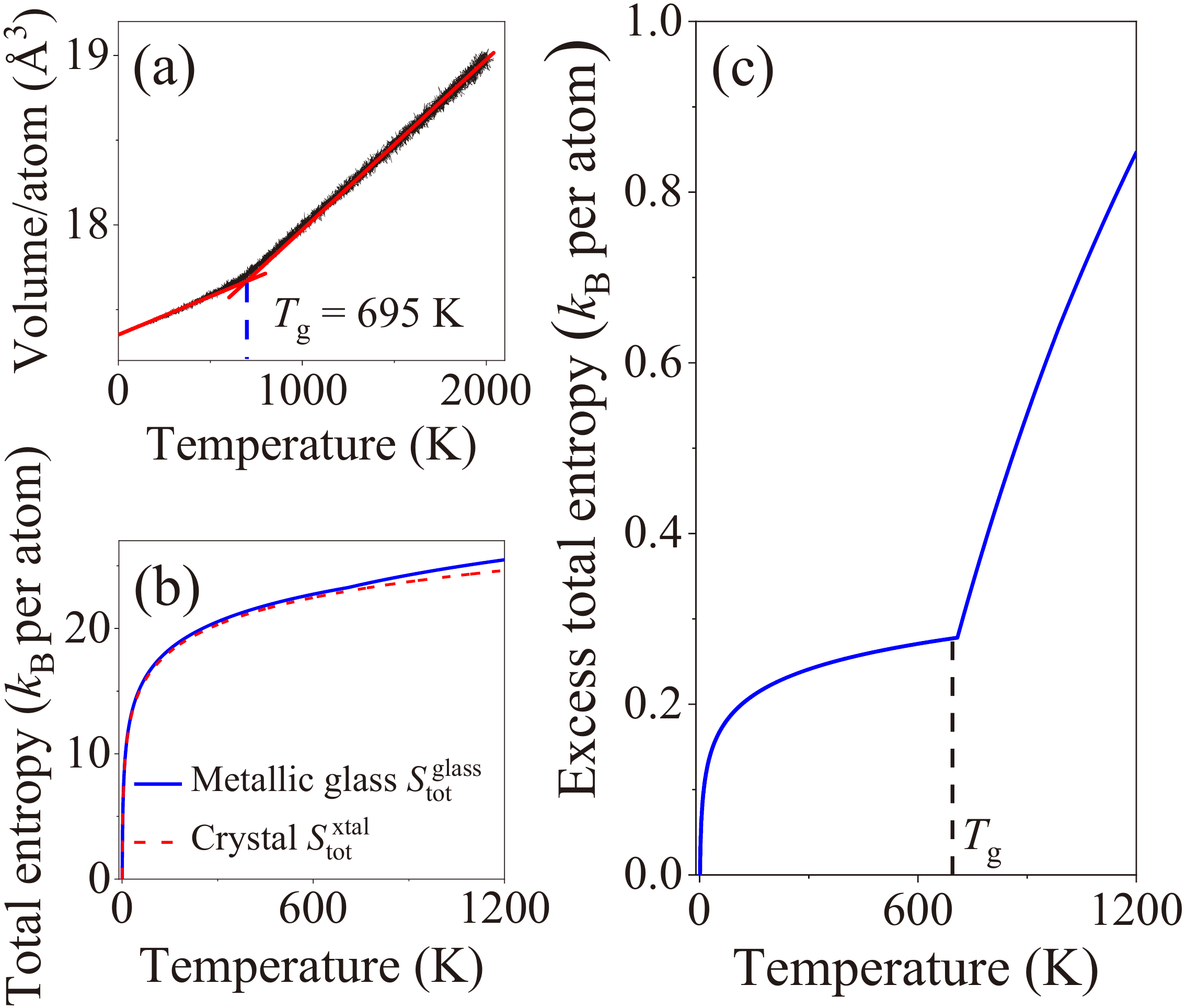}}
  \caption{Glass transition and excess total entropy.
  (a) Temperature dependence of the volume of a glass-forming liquid during cooling, the discontinuity in slope indicates the glass transition temperature, $T_{\rm g} = 695$ K.
  (b) Temperature dependence of total entropy in disordered phase and crystal.
  (c) Temperature dependence of excess total entropy $\Delta S_{\rm tot}$.
  }
  \label{fig1}
\end{figure}

Figure \ref{fig1} shows the variation of thermodynamic quantities as temperature involves. In Fig. \ref{fig1}(a), the glass transition is clearly signaled at $T_{\rm g} = 695$ K by a kink of the volume-temperature curve. The total entropy of glass and crystal over their 0 K reference is displayed in Fig. \ref{fig1}(b) from 0 K to 1200 K. One may notice that the total entropy diverges around glass transition. To further quantify this feature, we plot the excess total entropy, $\Delta S_{\rm tot}$, of glass and liquid over crystal in Fig. \ref{fig1}(c). $\Delta S_{\rm tot}$ is defined as
\begin{equation}\label{eq2}
\Delta S_{\rm tot} = S^{\rm glass}_{\rm tot} - S^{\rm xtal}_{\rm tot}.
\end{equation}
Here, the multiple crystalline phases are considered as references for the noncrystalline counterparts, i.e., the weighted mean of orthorhombic Cu$_{10}$Zr$_{7}$ and Laves CuZr$_{2}$ phases at 0 K-988.15 K, and a B2 Cu$_{50}$Zr$_{50}$ phase at 988.15 K-1208.15 K according to Cu-Zr binary phase diagram \cite{Arias1990}. In the following, the symbol $\Delta$ denotes the difference between glass (liquid) and crystal in a specific physical quantity. It is noted in Fig. \ref{fig1}(c) that the excess total entropy exhibits two kinks in the whole temperature range. The first one at low temperature is due to a change in vibrational entropy, which will be explained later in detail. With respect to the second kink at higher T, it coincides with the glass transition temperature where the excess entropy experiences a significant increase, consistent with the Adam--Gibbs entropic scenario \cite{adam1965temperature}.

\begin{figure}
\centering
 {\includegraphics[width=0.48\textwidth] {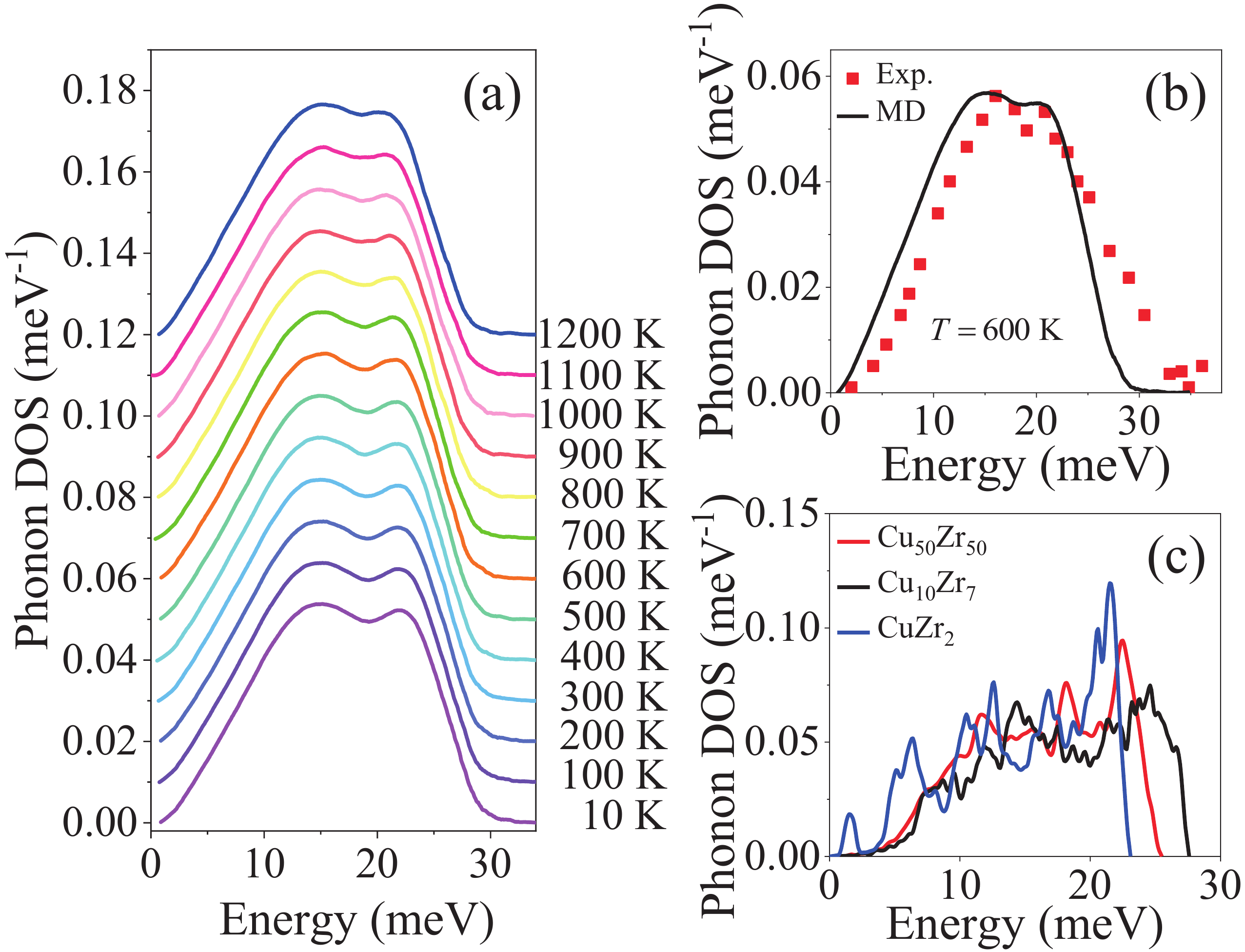}}
  \caption{Phonon of glass-forming liquid and crystal.
  (a) Phonon DOS of glass and liquid over wide temperature range.
  (b) Comparison of phonon DOS between simulation and experiment at 600 K.
  (c) Phonon DOS of three crystal phases including orthorhombic Cu$_{10}$Zr$_{7}$, Laves CuZr$_{2}$, and B2 Cu$_{50}$Zr$_{50}$.
  }
  \label{fig2}
\end{figure}

As for the vibrational entropy, we calculate the phonon DOS of both glass (liquid) and crystals as shown in Fig. \ref{fig2}. The phonon DOSs are calculated over a wide temperature range spanning from far below to  far above the glass transition. The data are displayed in Fig. \ref{fig2}(a) for selected temperatures from 10 K to 1200 K. To calibrate the simulations, we also compare the numerical data with INS experimental measurement of phonon at 600 K, as shown in Fig. \ref{fig2}(b). Although MD overestimates the soft modes and underestimates the high-frequency vibration, the simulations are overall comparable to the experiment. The extra soft modes in simulations are from the model preparation with an extremely high cooling rate due to the notoriously limited timescale in MD. For crystals, the vibrational DOSs of Cu$_{50}$Zr$_{50}$, Cu$_{10}$Zr$_7$ and CuZr$_2$ are all shown in Fig. \ref{fig2}(c), accounting for the different thermodynamically stable crystalline phases at different temperatures.

\begin{figure}
\centering
 {\includegraphics[width=0.48\textwidth] {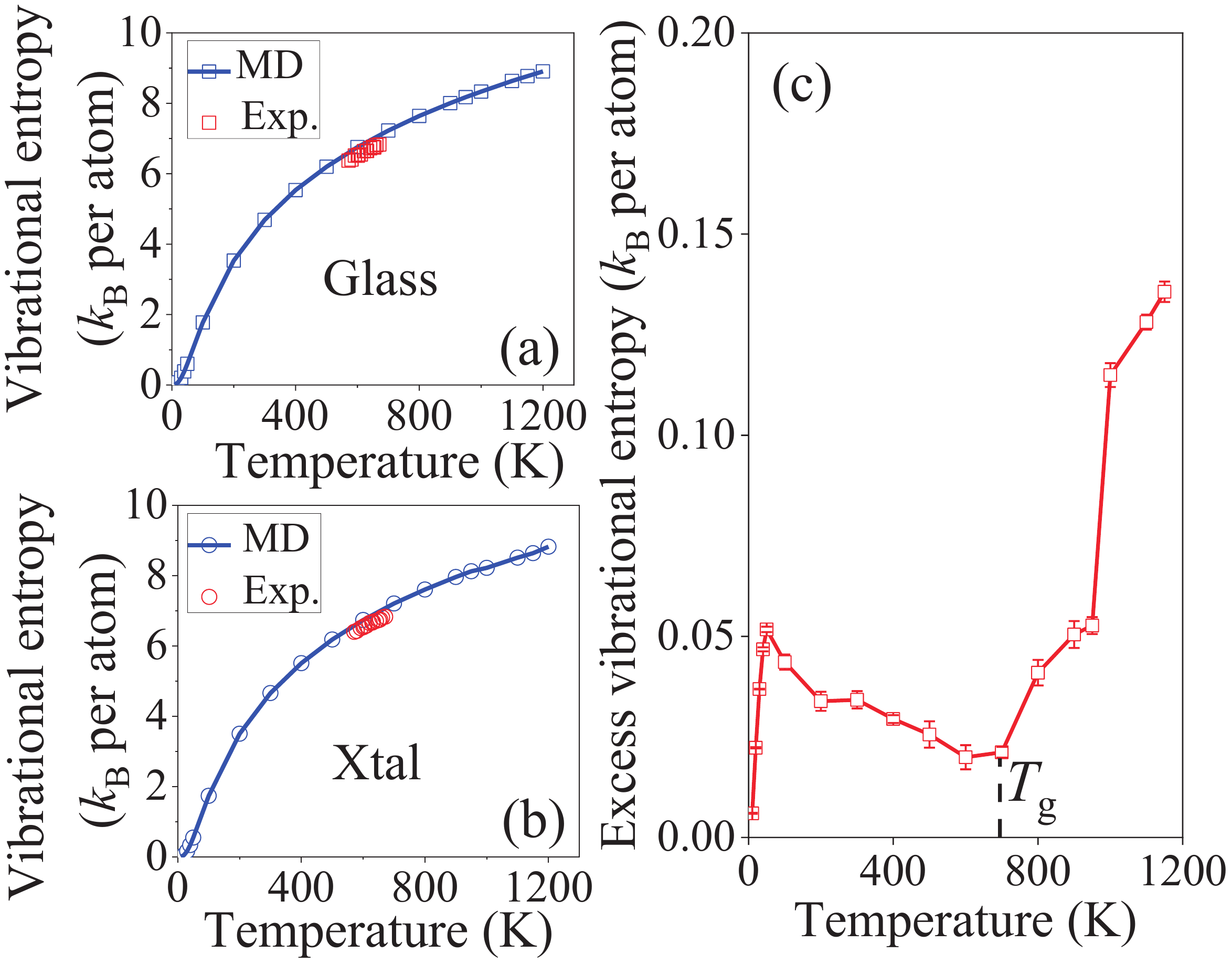}}
  \caption{Vibrational entropy of (a) glass-forming liquid, and (b) crystal as a function of temperature. The simulations are calibrated to the experiments. (c) Temperature dependence of excess vibrational entropy of glass and liquid over crystal. The error bars in (c) stand for the standard deviation of the entropy in five statistically independent configurations.}
  \label{fig3}
\end{figure}

Considering the bosonic nature of phonon, the vibrational entropy can be calculated from DOS via \cite{fultz2010vibrational,smith2017separating}
\begin{equation}\label{eq3}
\begin{aligned}
{S_{{\rm{vib}}}}\left( T \right) =& 3{k_{\rm{B}}}\int_0^\infty  {\rm{g}} \left( E \right)\left\{ \left[ {1 + n\left( T \right)} \right]\ln \left[ {1 + n\left( T \right)} \right]\right.\\
&\left.- {n\left( T \right)}\ln{n\left( T \right)} \right\}{\rm{d}}E.
\end{aligned}
\end{equation}
Here $n(T) =\{{\rm {exp}}[E/({k_{\rm B}}T)] - 1\}^{-1}$ is the Bose-Einstein occupation number with $k_{\rm B}$ the Boltzmann constant. ${\rm g}(E)$ is the normalized phonon DOS and $E = \hbar \omega_l$ is the phonon energy. The calculated vibrational entropy versus temperature is shown in Fig. \ref{fig3}(a-b) for both glass and crystals. It is seen that our calculations agree quantitatively with experiments. The vibrational entropies evolve continuously from glass to liquid without any noticeable discontinuity. The excess vibrational entropy of glass (liquid) over crystal $\Delta S_{\rm vib} = S^{\rm glass}_{\rm vib} - S^{\rm xtal}_{\rm vib}$ is further plotted in Fig. \ref{fig3}(c), which exhibits two kinks in analogy with the excess total entropy; see Fig. \ref{fig1}(c).

\begin{figure}
\centering
 {\includegraphics[width=0.45\textwidth] {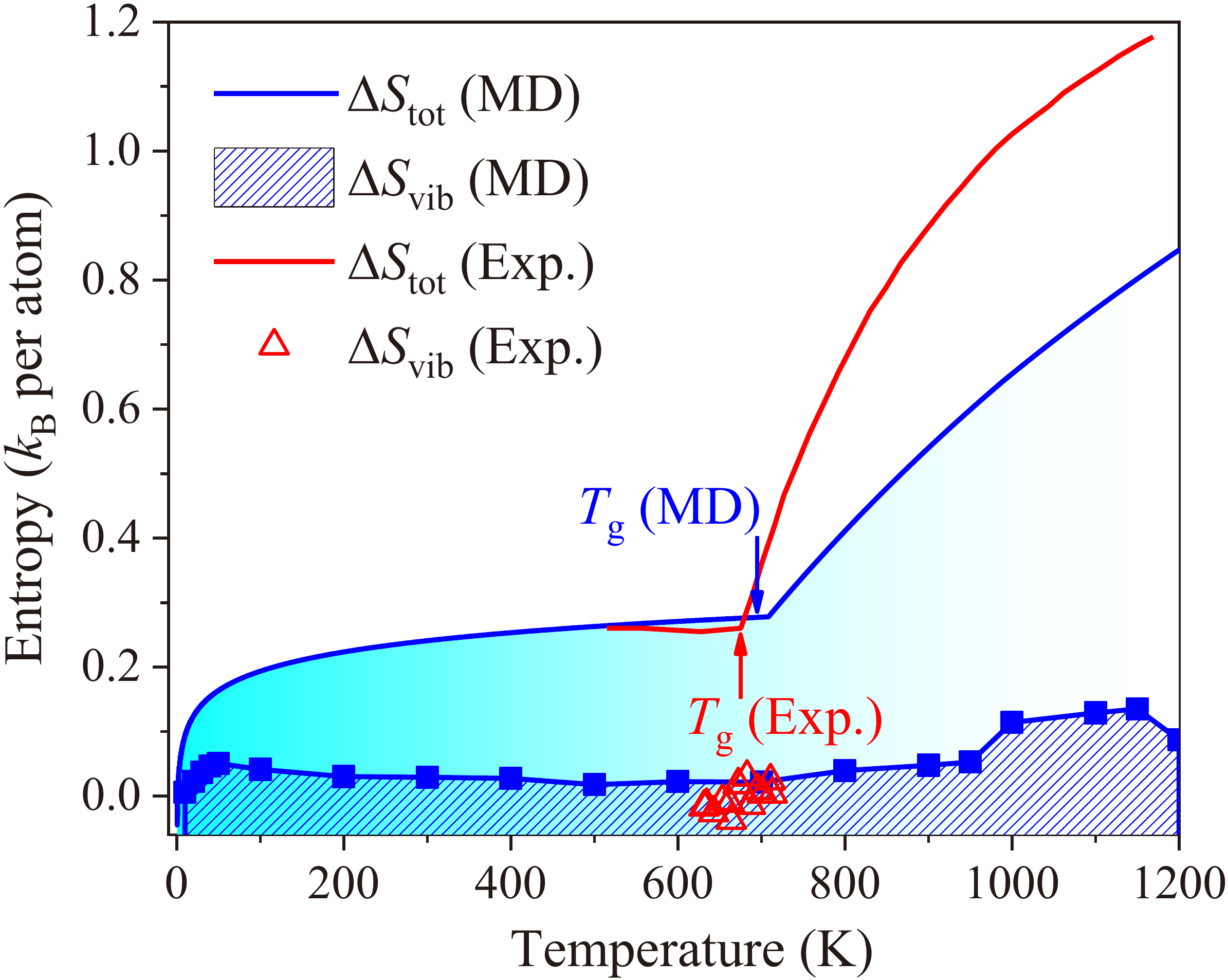}}
  \caption{Panorama of excess total and vibrational entropy over entire temperature range. The excess total entropy experiences abrupt increase upon glass transition with dominating contribution from its configurational component. The simulations are calibrated by experiments.}
  \label{fig4}
\end{figure}

\begin{figure*}
\centering
 {\includegraphics[width=1\textwidth] {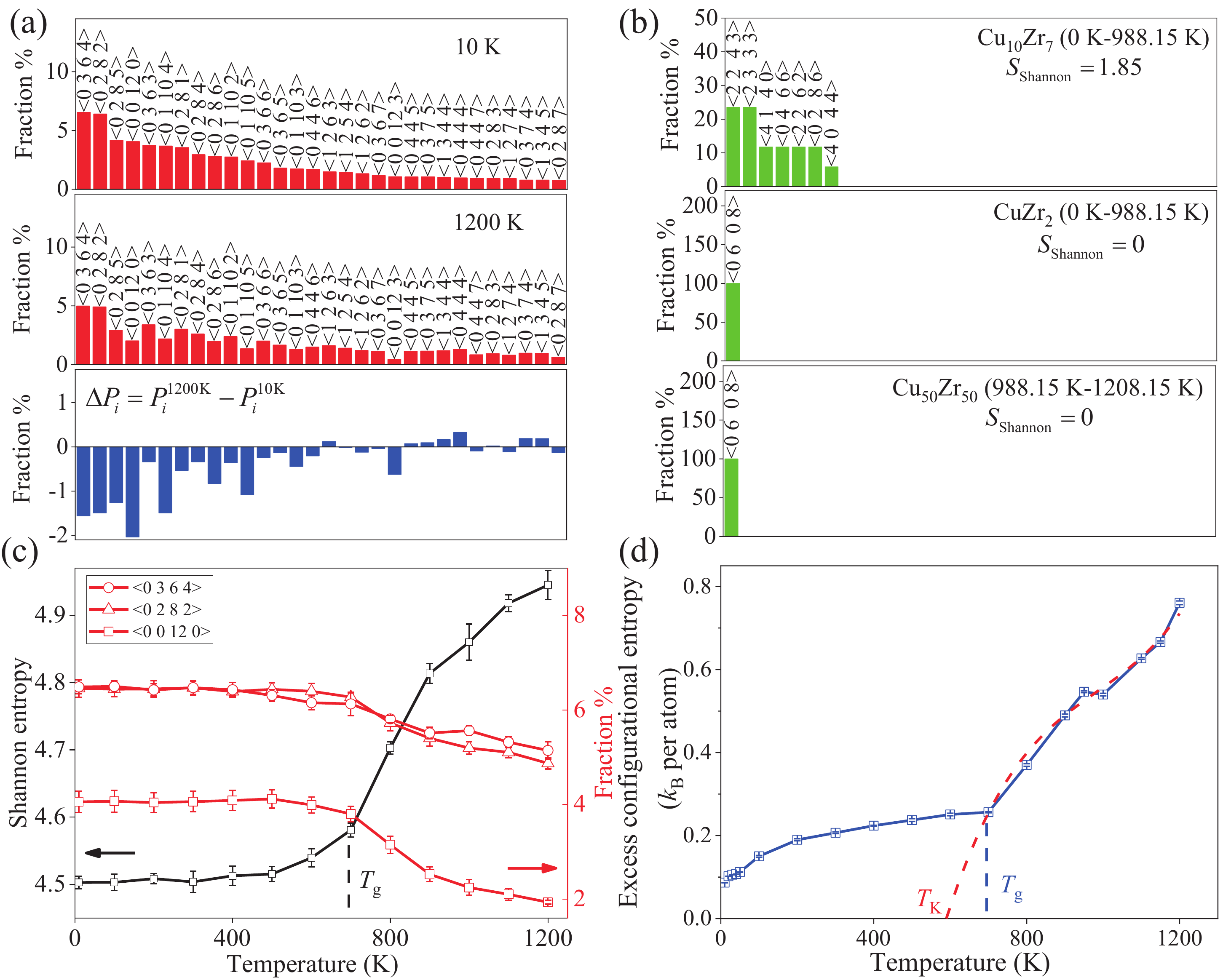}}
  \caption{Shannon entropy of local structures across the glass transition as a microscopic measure of configurational entropy.
  (a) Distribution and variation of 30 most frequent Voronoi polyhedra at 10 K and 1200 K, respectively.
  (b) Distribution of Voronoi polyhedra in crystals.
  (c) Temperature dependence of Shannon entropy which is extracted from the distribution of local structures. The fractions of specific Voronoi polyhedra $\left\langle 0, 3, 6, 4 \right\rangle$, $\left\langle 0, 0, 12, 0 \right\rangle$ and $\left\langle 0, 2, 8, 2 \right\rangle$ are also shown for comparison. Each data point is an average of five independent inherent structures, with error bars in (c) denoting the standard deviation of entropy in five statistically independent configurations.
  (d) Temperature dependence of excess configurational entropy of glass-forming liquid, which is extracted from the difference between total entropy and the vibrational entropy as plotted in Fig. \ref{fig4}. The error bars in (d) stand for the standard deviation of entropy in five statistically independent configurations.
  }
  \label{fig5}
\end{figure*}

Once one has the excess total entropy $\Delta S_{\rm tot}$ and the excess vibrational entropy $\Delta S_{\rm vib}$, the exact role of configurational entropy
\begin{equation}\label{eq4}
\Delta S_{\rm conf} = \Delta S_{\rm tot} - \Delta S_{\rm vib}
\end{equation}
played in glass transition can be examined by subtracting the vibrational part from the total entropy. The excess entropy of glass (liquid) over crystal is summarized in Fig. \ref{fig4} over the entire temperature range, which includes the temperature observation window of the experiments. The excess vibrational entropy is trivial in most of the temperature range compared with configurational entropy. The striking variation in excess entropy near the glass transition is mainly from the change in configurational entropy, whereas the vibrational entropy varies moderately at the transition. The bare configurational entropy is further shown in Fig. \ref{fig5}(d) to clearly confirm its critical role in the glass transition. Such data unambiguously support the Adam--Gibbs entropy scenario and are consistent with experimental observations \cite{smith2017separating}.

However, as a special case, the relative contribution of configurational entropy and vibrational entropy to the excess total entropy may be comparable, if only at extremely low temperature near 0 K. In this condition, the vibrational entropy is more sensitive to temperature and should increase markedly from zero to a finite value at the very beginning of heating from 0 K; see Fig. 3(c). Physically, the entropy from dynamical sources increases as temperature goes up because the system explores a larger volume in the phase space with stronger excitations of dynamical degrees of freedom \cite{fultz2010vibrational}. Finally, we note that $T_{\rm g}$ in simulations is a bit higher than that found in experiments. The slight difference is understandable since the MD model is being quenched much faster,  which makes the inherent structure remain on higher positions in the PEL.

Now that we have tested the validity of the Adam--Gibbs scenario, the remaining unsolved issue is whether there is any unambiguous structural variation that is linked with the evolution of the configurational entropy. This link is crucial to explain the variation in linear response to external fields such as shear, which is deeply rooted in the microstructure \cite{Cubuk2017}. However, if one cares only about the structure at the level of two-body correlation, or the fraction of a specific local structure, usually there is no dramatic change accounting for the dynamical arrest. To figure out the hidden variables, the Voronoi tessellation scheme is adopted to analyze the inherent structures \cite{Sheng2006,Cheng2010}.

The distributions of the 30 most frequent Voronoi polyhedra are displayed in Fig. \ref{fig5}(a) for both glass and liquid. Although the geometries of the clusters do not change much from liquid to glass, their distribution does change as evidenced by the difference, $\Delta P_i$, of fractions at 1200 K and 10 K, respectively. The geometric structures that are most frequent in the deep glass state become less populated in the liquid state upon crossing the glass transition. Consequently, the local structures in the liquid are more evenly distributed, which in turn increases the diversity of structures and the corresponding configurational entropy. In order to compare with disordered states, we also list the local structures of the three crystalline phases in Fig. \ref{fig5}(b). Only very few local structures are present in crystals, indicating very low configurational entropy.

In order to further quantify the diversity of structures and the configurational entropy, we introduce the concept of Shannon information entropy \cite{shannon1948mathematical,Briscoe2008,Yoon2019a} associated with the incidence of local Voronoi structures (polyhedra) in disordered liquid and glass states \cite{Wei2019JCP}, which reads
\begin{equation}\label{eq5}
{S_{{\rm{Shannon}}}} =  - \sum\limits_{i = 1}^n {{P_i}\left( {{x_i}} \right)\ln {P_i}\left( {{x_i}} \right)},
\end{equation}
here $P(x_i)$ is the normalized probability density of a Voronoi polyhedron $x_i$. The Shannon entropy provides a ``solid and quantitative basis for the interpretation of the thermodynamic entropy''~\cite{Ben-Naim}, and here we use it as a qualitative measure for the evolution of the configurational entropy in the physical system.

\begin{figure}
\centering
 {\includegraphics[width=0.40\textwidth] {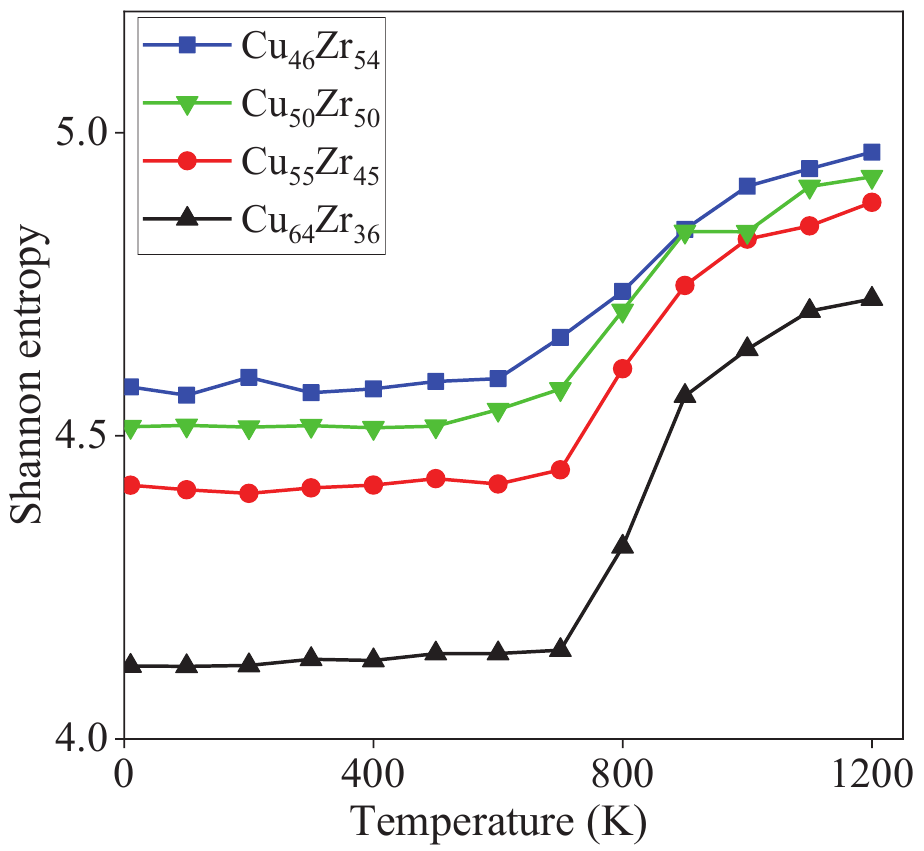}}
  \caption{Composition dependence of the Shannon entropy about Voronoi structures. The data of four different metallic glass systems£¬i.e., $\mathrm{Cu}_{46} \mathrm{Zr}_{54}, \mathrm{Cu}_{55} \mathrm{Zr}_{45}, \mathrm{Cu}_{64} \mathrm{Zr}_{36}$ and $\mathrm{Cu} 50 \mathrm{Zrso}$, include 32000 atoms, are listed for comparison. Even if the composition varies, the remarkedly change of Shannon entropy on glass transition is clearly seen and is general for all the compositions.}
  \label{fig6}
\end{figure}

The computed the Shannon entropy of glass-forming systems is shown in Fig. \ref{fig5}(c). It is seen that the Shannon entropy rises abruptly at the glass transition temperature, which is signaled by a kink. As Fig. \ref{fig5}(d) demonstrates, the variation of excess configurational entropy $\Delta S_{\rm conf}$ (Eq. (\ref{eq4})) is in very good qualitative agreement with the change in Shannon entropy $\Delta S_{\rm Shannon}$. This is not a coincidence but a strong evidence that the change of diversity in local atomic structures yields an abrupt rise of configurational entropy as suggested in the Adam--Gibbs entropic scenario. It is clear that Shannon entropy is a universal composition-independent metric to capture the glass transition, as shown in Fig. \ref{fig6}.

\begin{figure}
\centering
 {\includegraphics[width=0.40\textwidth] {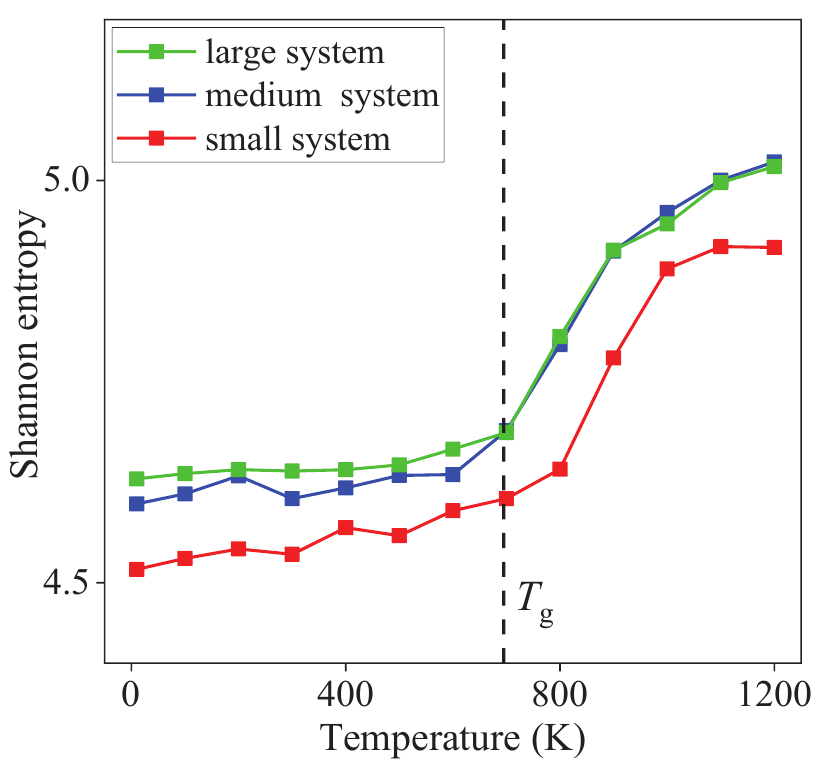}}
  \caption{Size dependence of the Shannon entropy about Voronoi structures. Three independent large, medium and small systems include 31250, 21296 and 10976 atoms, respectively. As the size becomes larger, the Shannon entropy is bigger with increase in structural diversity. But the data of Shannon entropy convergence to a fixed value if the system size is approximately of the large system. Therefore, the present model with 31,250 atoms is believed to yield reliable statistics on Voronoi polyhedra.}
  \label{fig7}
\end{figure}

However, if one examines the temperature dependence of specific local structures, such as $\left\langle 0, 3, 6, 4 \right\rangle$, $\left\langle 0, 0, 12, 0 \right\rangle$ and $\left\langle 0, 2, 8, 2 \right\rangle$, as shown in Fig. \ref{fig5}(c), only $\left\langle 0, 0, 12, 0 \right\rangle$ changes pronouncedly across glass transition temperature. Note that none of the Voronoi structures has a fraction larger than 5\%, which means there is no dominating structures in the glass former; thus the change in any specific structure may not be able to reflect complete structural information about configurational variation \cite{Cheng2008}. Therefore, Shannon entropy presents more relevant statistical information about the whole scenario of structural evolution across the glass transition. Thus, we do find a hidden structural change across the glass transition from a statistical perspective, which has been puzzling for decades. For a first approximation, we deduce the Kauzmann temperature as $T_{\rm K} = 590$ K by extrapolating the excess configurational entropy to zero, which is supported by the experimental data, such as $T_{\rm K} = 571$ K in Ref. \cite{Wang2014}, and $T_{\rm K} = 627$ K in Ref. \cite{smith2017separating}. Finally, we note that a larger model with 31,250 atoms is applied for the Voronoi structure analysis. As seen in Fig. \ref{fig7}, such model size yields nearly converged diversity of local Voronoi structures.

We further propose to use a centrosymmetry parameter to characterize the feature of local atomic environment in glass and liquid, since the parameter can be an effective measure of the degree of local disorder around an atom. It is usually used to recognize versatile structural defect in crystals. For each atom, it is defined as follows:
\begin{equation}\label{eq6}
P=\sum_{i=1, n}\left|\vec{R}_{i}+\vec{R}_{i+n}\right|^{2},
\end{equation}
where $\vec{R}_{i}$ and $\vec{R}_{i+n}$ are the vectors or bonds representing the $n$ pairs of opposite nearest neighbors of the atom. By adding each pair of vectors together, the sum of the squares of the $n$ resulting vectors, i.e., the centrosymmetry parameter is derived. It is a scalar. The magnitude of the centrosymmetry parameter provides a metric of the departure from centrosymmetry in the immediate vicinity of any atom considered. The higher the centrosymmetry parameter is, the stronger the structural disorder and the more non-centrosymmetric the local atomic environment.

\begin{figure}
\centering
 {\includegraphics[width=0.40\textwidth] {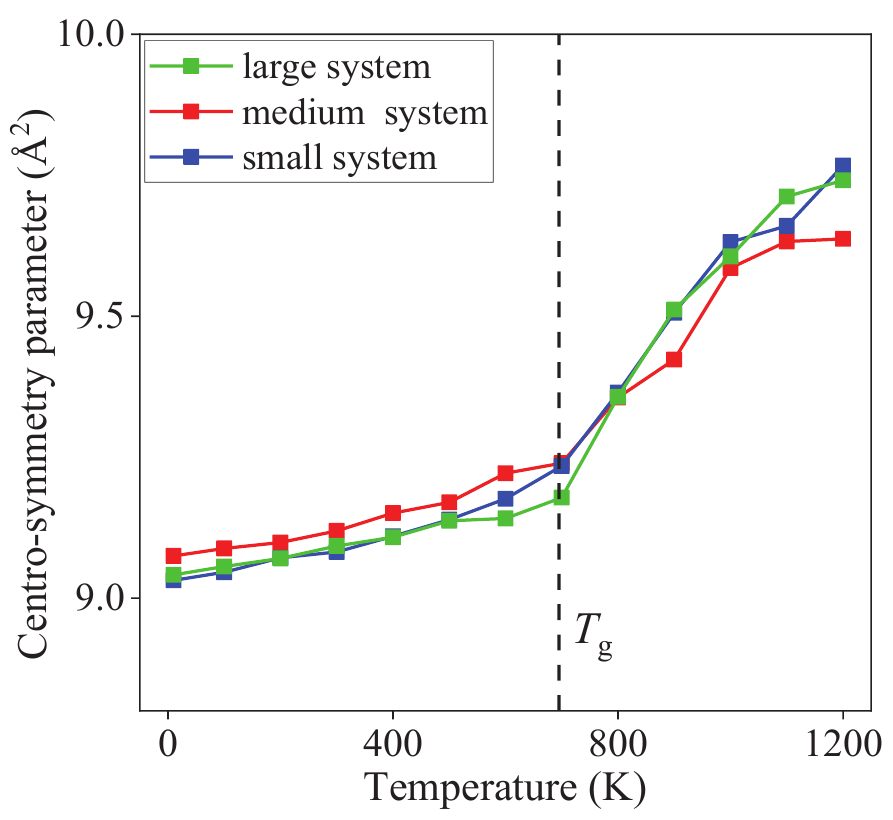}}
  \caption{Ensemble average of the centrosymmetry parameter for systems with different number of atoms. As the size becomes larger, the centro-symmetry parameter changes slightly (the kink becomes more pronounced) with the increase of structural diversity. There is no clear size effect on this parameter, compared with Shannon entropy as shown in Fig. \ref{fig7}.}
  \label{fig8}
\end{figure}

It is evident that the increase of the fraction of \textit{certain} geometrically favoured local structures upon lowering the temperature is linked to the reduced degree of local centrosymmetry \cite{Milkus2016}, as demonstrated by the evolution of the average non-centro-symmetry parameter, as shown in Fig. \ref{fig8}. This fact reflects the fact that certain structures (Voronoi polyhedral) which are centrosymmetric (e.g. dodecahedra with $\langle 0,0,12,0 \rangle$) become more frequent as temperature decreases. This brings along an increase of rigidity, because centrosymmetric structures do not possess nonaffine softening modes, as demonstrated in \cite{Milkus2016}. The proliferation of these centrosymmetric structures is fast with decreasing $T$ in the liquid phase, until a situation is reached where rigidity is such that further structural adjustments become energetically unfavorable and the system gets frozen-in at the glass transition. Upon further decreasing T in the solid glass the configuration entropy thus flattens out due to high penalty for structural rearrangements caused by rigidity. This mechanism provides an explanation for the kink which signals the glass transition in the configurational entropy. Hence the configurationally favored structures give higher rigidity and lower boson peak (lower soft modes) \cite{Milkus2016,Yang2019}. In terms of both Voronoi polyhedra and entropies, a natural link between entropy and rigidity/elasticity of glass-forming systems can thus be established. Therefore, the present results may lead to a unification of apparently different concepts of glass transition, i.e., entropy in the Adam--Gibbs sense \cite{adam1965temperature}, the shear modulus in the shoving-model \cite{Dyre2006,Dyre2008,dyre2009brief,Krausser2015} and Frenkel's viscoelastic crossover~\cite{Trachenko2013,Trachenko2016,Yoon2019b}.

\section{Conclusion}

Our results provide microscopic insights into the Adam--Gibbs entropic scenario of the glass transition in a model atomic glass former via quantification of temperature-dependent total, vibrational, and configurational entropies. The change of entropy that dominates the glass transition is confirmed to be originated mostly from configurational entropy, while the vibrational entropy is featureless at the transition. The findings are in agreement with recent INS experiments~\cite{smith2017separating}, and independent simulations~\cite{Antonelli} and provide additional atomistic details. The hidden emergence of atomic-level structures leading to dynamical arrest is unambiguously revealed by studying the distributions of local Voronoi polyhedra in terms of Shannon information entropy. In particular, upon decreasing $T$ in the liquid it is seen that a limited number of Voronoi polyhedra become more frequent with respect to all the others, which makes the distribution of Voronoi polyhedra more uneven and thus reduces the configurational and Shannon entropies. Since the favoured polyhedra are associated with a higher degree of local centrosymmetry, hence with mechanical rigidity (they have less nonaffine softening modes~\cite{Milkus2016}), this eventually leads to a rigidification process at the glass transition, after which further structural rearrangements become energetically unfavourable due to the rigid environment, and the configurational entropy then decreases much less with further decreasing $T$ in the solid glass.

The emerging scenario may pave the way for constructing a complete framework eventually connecting structure, entropy and viscoelasticity at the glass transition of liquids.

\section*{Acknowledgement}

The idea of using Shannon information entropy in analyzing the diversity of local structures is stimulated by the discussions with Jeppe Dyre and Peter Harrowell. Useful discussions with Stefano Martiniani are also gratefully acknowledged. This work is supported by the National Key Research and Development Program of China (No. 2017YFB0701502 and No. 2017YFB0702003), and the NSFC (No. 11672299 and No. 11790292). The Key Research Program of Frontier Sciences (No. QYZDJSSW-JSC011) and the Youth Innovation Promotion Association of Chinese Academy of Sciences (No. 2017025) are also acknowledged.

%

\end{document}